\documentstyle[11pt,newpasp,psfig]{article}
\begin{document}

\title{Formation of Gaseous Shells }
\author{F. Combes and V. Charmandaris}
\affil{DEMIRM, Observatoire de Paris, 61 Av. de l'Observatoire,
F-75 014, Paris, France}

\begin{abstract}
HI observations have revealed in several shell galaxies the presence
of gaseous shells slightly displaced from the stellar shells radially,
in the outward direction. We propose a mechanism to form this gaseous
shells, based on the well-known phase-wrapping process of the
companion matter in a merger, with nearly radial orbits. The mechanism
relies on the existence of a clumpy interstellar matter, and on
dynamical friction experienced by the companion core.
\end{abstract}


\vspace*{-0.5cm}
\section{The Problem}
\vspace*{-0.2cm}

Shells are sharp-edged features, formed during interactions and
mergers, through phase-wrapping of debris (Quinn 1984, Dupraz \&
Combes 1986, Hernquist \& Quinn 1989).  Recent HI observations have
revealed the existence of associated gaseous shells, slightly
displaced from the stellar ones, questioning the validity of the
phase-wrapping mechanism (Centaurus A: Schiminovich et al 1994; NGC
2865: Schiminovich et al 1995; NGC 1210: Petric et al 1997).  An
intriguing result is that the HI shells follow the curvature of the
stellar shells, but are shifted about 1kpc outside.

There are two ways shells can be formed: {\bf -(1)-} in minor mergers,
shells correspond to phase-wrapping of the stars liberated from the
small companion (e.g. Quinn 1984); {\bf -(2)-} in major mergers,
shells correspond to phase-wrapping of the debris falling back into
the merged-object potential (Hernquist \& Spergel 1992, Hibbard \&
Mihos 1995). But what is the fate of gas? due to dissipation, it falls towards the
center, as in the simulations of Weil \& Hernquist (1993), and
 there should not exist gaseous shells.

\vspace*{-0.5cm}
\section{Solution Proposed}
\vspace*{-0.2cm}

There exists a population of small and dense gas clouds, that have
very low dissipation.  This gas has a behaviour intermediate between
stars and diffuse gas, and remains available to form shells.  
Already Kojima \& Noguchi (1997) have simulated the sinking of
a disk satellite into an elliptical, with a sticky particle code, instead of
SPH, and found no segregation between gas and stars. We have also
simulated the phenomenon, with a cloud-cloud collision code, to be
able to control the dissipation rate.

\begin{figure}[t]
\centerline{
\psfig{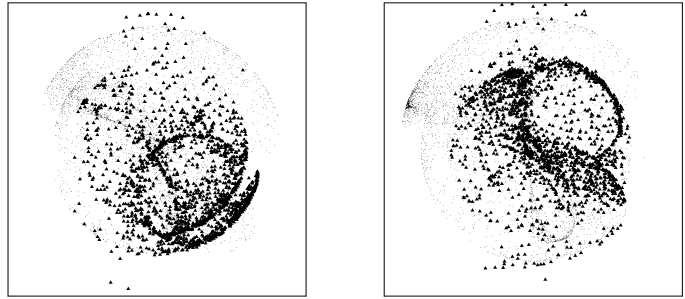}
}
\caption{Simulations of the infall of a small spiral satellite into a 
giant elliptical, with low dissipation gas, and dynamical friction
taken into account. The stars are black triangles, and 
gas clouds, dots.} 
\label{fig1}
\end{figure}

With strong dissipation, the gas component, after a few oscillations
back and forth in the primary's potential, settles in the center, as
previously. But with small dissipation, only a small fraction of the
gas falls into the potential well, most of it form shells (cf figure
\ref{fig1}).

Now it is necessary to explain the spatial displacement between the
gaseous and stellar shells.  Two possibilities could be tested: {\bf
-(1)-} the gas in the companion is not as bound, and does not occupy
the same region initially, being in the outer parts of the galaxy.  We
simulate this, but this results in only a very slight and negligible
shift.  {\bf -(2)-} in the merging, the gas is liberated early from
the companion by the tidal forces, since it is not very bound, while
the stars are liberated afterwards.  Through dynamical friction, the
stars have therefore time to lose a lot of energy, contrary to the
gas.  Dynamical friction also explains the dynamical range of the
shell radii (Dupraz \& Combes 1987).

 This second possibility accounts very well for the shift observed
between HI and stellar shells, according to simulations.

\vspace*{-0.5cm}
\section{Observations}
\vspace*{-0.2cm}

To check the model, millimeter observations have been carried out, to
detect molecular gas in shells, since the dense clumps able to form
shells should be on H$_2$ form.  This led to the surprising detection
of CO with the SEST 15m-telescope in Centaurus-A shells (Charmandaris,
Combes, van der Hulst 1999).  There are comparable amounts of H$_2$
and HI gas in the shells, far away from the galaxy center (15 kpc),
which is completely unusual for a normal galaxy.  This is compatible
with the view that the dense clumps have been dragged in the shells by
the phase-wrapping mechanism, and the HI diffuse gas has been photo-
dissociated from there.

\end{document}